\documentclass[journal]{IEEEtran}
\ifCLASSINFOpdf
\else
   \usepackage[dvips]{graphicx}
\fi
\usepackage{url}

\usepackage{graphicx}
\usepackage{subfigure}

\begin{document}

\title{Decorrelated Unbiased Converted Measurement for Bistatic Radar Tracking}

\author{Sen Wang, Qinglong Bao
\thanks{Manuscript received March 29, 2020.}
\thanks{The authors are with National Key Laboratory of Science and Technology on ATR, National University of Defense Technology, Changsha, China (e-mail: wangsen11@nudt.edu.cn, baoqinglong@nudt.edu.cn)}}

\markboth{}
{Shell \MakeLowercase{\textit{et al.}}: Bare Demo of IEEEtran.cls for IEEE Journals}
\maketitle

\begin{abstract}
Tracking with bistatic radar measurements is challenging due to the fact that the measurements are nonlinear functions of the Cartesian state. The converted measurement Kalman filter (CMKF) converts the raw measurement into Cartesian coordinates prior to tracking, which avoids the pitfalls of the extended Kalman filter (EKF). The challenges of CMKF are debiasing the converted measurement and approximating the converted measurement error covariance. Due to no closed form of biases, this letter utilizes the second order Taylor series expansion of the conventional measurement conversion to find the conversion bias in bistatic radar, which derives the Unbiased Converted Measurement (UCM). In order to decorrelate the converted measurement error covariance from the measurement noise, the prediction is utilized to evaluate the covariance, which derives the Decorrelated Unbiased Converted Measurement (DUCM). Monte Carlo simulations show that the DUCM is unbiased and consistent, and the DUCM filter exhibits the improved performance compared with the conventional CMKF and the UCM filter in bistatic radar tracking.
\end{abstract}

\begin{IEEEkeywords}
Decorrelated unbiased converted measurement, converted measurement Kalman filter, bistatic radar tracking.
\end{IEEEkeywords}

\IEEEpeerreviewmaketitle

\section{Introduction}

\IEEEPARstart{T}{racking} with bistatic radar measurements is confronted with the challenge that the state dynamic equation is linear whereas the measurements are nonlinear functions of the Cartesian state. Typically, some approaches to deal with nonlinear measurements convert the state prediction into the measurement coordinates using the nonlinear observation function, such as the extended Kalman filter (EKF), the unscented Kalman filter (UKF) \cite{PETSIOS2007665} and the cubature Kalman filter (CKF) \cite{4982682}.

The converted measurement Kalman filter (CMKF) converts the raw measurement into Cartesian coordinates prior to tracking. Consequently, the converted measurement is a linear function of the state, allowing for the use of a standard (linear) Kalman filter. The CMKF with a bias correction term has previously been shown to have superior performance over the converted prediction approach \cite{Lerro1993}. \cite{4646978} performs elliptic-to-Cartesian coordinate conversion of the measurements and evaluates the noise affecting Cartesian coordinate by a non-diagonal matrix. \cite{7943869} approximates the non-Gaussian Cartesian error distribution by a single Gaussian distribution with first two moments matched using first and second order Taylor expansion of the conversion, and fits a Gaussian mixture distribution model to the converted measurement. Converted measurement sigma point Kalman filter for bistatic sonar and radar tracking \cite{Bordonaro2019} uses a sigma point transform to estimate the conversion bias and the converted measurement error covariance, which do not need to be explicitly derived. Analogously, various approaches for bistatic to Cartesian conversion using cubature points were proposed, and were compared in \cite{Crouse2014}. The only known works that directly address the bias issue in bistatic coordinate conversion are \cite{Coogle2012}, which gives the debiased conversion, and \cite{Coogle2013}, which uses the second-order unscented transform. The challenges of CMKF are debiasing the converted measurement and approximating the converted measurement error covariance. Compared with the monostatic radar \cite{Bishop2007,Bordonaro2014,6809917,8060575}, these two problems in the bistatic case have not been addressed well.

In this letter, based on Taylor series expansion, the Unbiased Converted Measurement (UCM) for bistatic measurement conversion is proposed. In addition, the Decorrelated Unbiased Converted Measurement (DUCM) evaluates the converted measurement error covariance using the prediction, which decorrelates the covariance from the measurement noise. Simulation results show that the DUCM has the best performance in both static conversion and dynamic tracking compared with the conventional conversion and the UCM.

\section{Background}

The geometry of bistatic radar system is described in Fig. 1. For simplicity, we assume that the target \emph{Tgt}, transmitter \emph{Tx}, and receiver \emph{Rx} are in the same plane. Without loss of generality, the receiver is assumed to be at the origin and the transmitter is at $[L,0]^T$ in Cartesian coordinates, where the baseline $L$ is available. The position of target \emph{Tgt} can be described as $t=[x,y]^T$ in Cartesian coordinates, or as $z=[b,\alpha]^T$ in measurement coordinates, where the bistatic range $b=r_1+r_2$ is the total range from transmitter to target to receiver, and $\alpha$ is the bearing to the target. The relationship between Cartesian coordinates and measurement coordinates is as follows:
\begin{equation}
z = h\left( t \right) = \left[ {\begin{array}{*{20}{c}}
{\varphi \left( {x,y} \right)}\\
{\gamma \left( {x,y} \right)}
\end{array}} \right] = \left[ {\begin{array}{*{20}{c}}
{\sqrt {{x^2} + {y^2}}  + \sqrt {{{\left( {x - L} \right)}^2} + {y^2}} }\\
{\arctan \left( {{y \mathord{\left/
 {\vphantom {y x}} \right.
 \kern-\nulldelimiterspace} x}} \right)}
\end{array}} \right]
\end{equation}
\begin{equation}
\label{E1}
t=h^{-1}(z)=\left[
\begin{array}{c}
f(b,\alpha)\\
g(b,\alpha)
\end{array}
\right]
=\left[
\begin{array}{c}
\frac{(L^2-b^2)\cos\alpha}{2(L\cos\alpha-b)}\\
\frac{(L^2-b^2)\sin\alpha}{2(L\cos\alpha-b)}
\end{array}
\right]
\end{equation}
\begin{figure}
\centerline{\includegraphics[width=7cm]{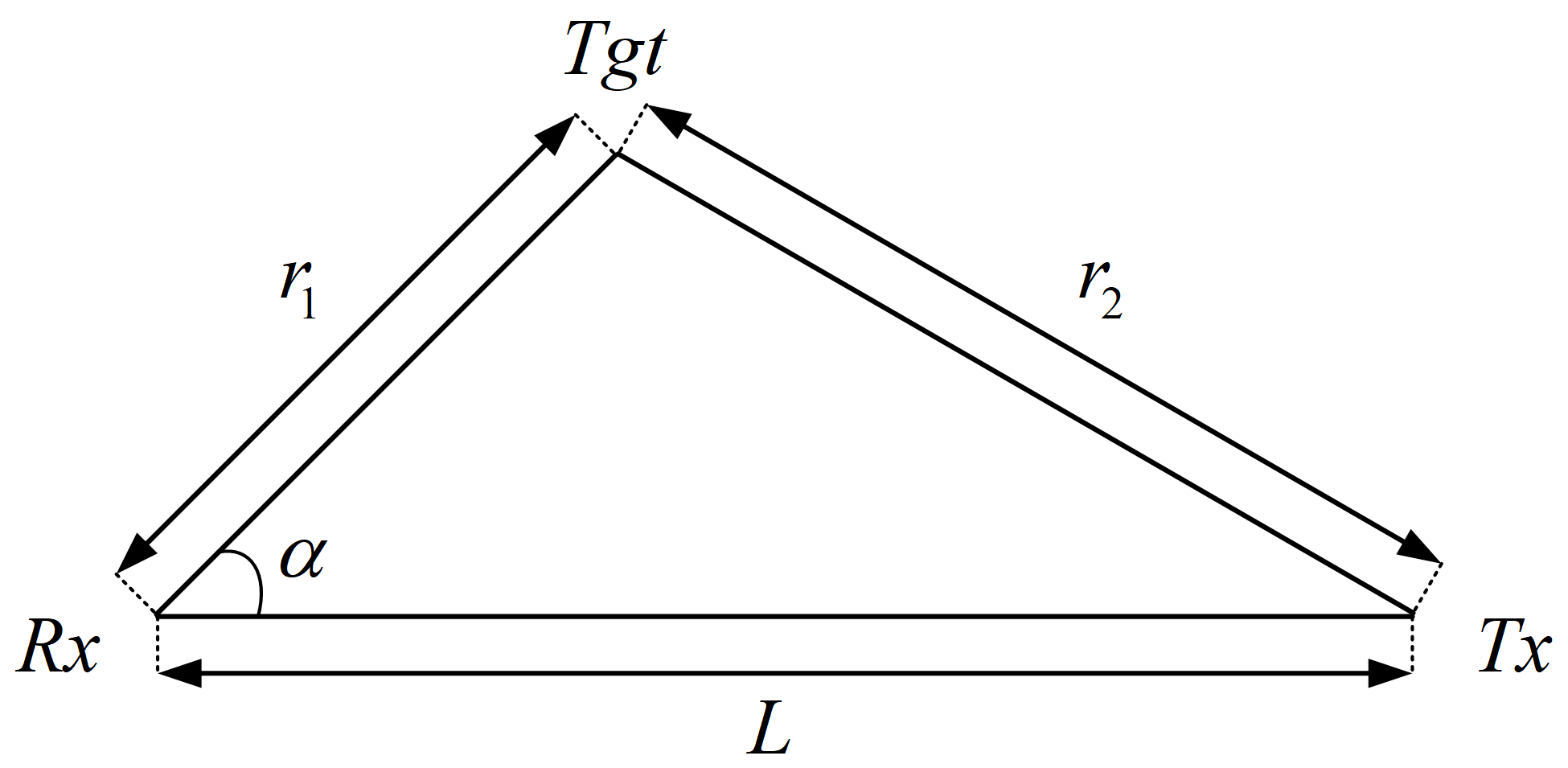}}
\caption{Geometry of the target \emph{Tgt}, transmitter \emph{Tx}, and receiver \emph{Rx} of bistatic radar system.}
\end{figure}
The raw measurement of the position of a target is obtained in measurement coordinates as
\begin{equation}
\begin{array}{l}
{b_m} = \bar b + {w_b}\\
{\alpha _m} = \bar \alpha  + {w_\alpha }
\end{array}
\end{equation}
where $\bar b$ and $\bar \alpha$ are the ground truth of bistatic range and bearing, ${w_b}$ and ${w_\alpha}$ are independent zero-mean Gaussian white noise with variance ${\sigma _b^2}$ and ${\sigma _\alpha ^2}$, respectively. Consequently, the PDF of measurement ${z_m} = {\left[ {{b_m},{\alpha _m}} \right]^T}$ is
\begin{equation}
{p_{b\alpha }}\left( {{z_m}} \right) = \frac{1}{{2\pi \sqrt {\det \left( {{R_{b\alpha }}} \right)} }}\exp \left( { - \frac{1}{2}{{\left( {{z_m} - \bar z} \right)}^T}R_{b\alpha }^{ - 1}\left( {{z_m} - \bar z} \right)} \right)
\end{equation}
where $\bar z = {\left[ {\bar b,\bar \alpha } \right]^T}$ and ${R_{b\alpha }} = diag\left( {\sigma _b^2,\sigma _\alpha ^2} \right)$. If the point ${z_m}$ is converted into Cartesian coordinates as ${t_m} = {h^{ - 1}}\left( {{z_m}} \right)$, then the converted PDF was given by \cite{papoulis2002probability}
\begin{equation}
{p_{xy}}\left( {{t_m}} \right) = \frac{{{p_{b\alpha }}\left( {h\left( {{t_m}} \right)} \right)}}{{\left|
J
\right|}}
\end{equation}
where $\left|J\right|$ represents the determinant of the Jacobian matrix, which is defined to be
\begin{equation}
J = \left[ {\begin{array}{*{20}{c}}
{\frac{{\partial f}}{{\partial b}}}&{\frac{{\partial f}}{{\partial \alpha }}}\\
{\frac{{\partial g}}{{\partial b}}}&{\frac{{\partial g}}{{\partial \alpha }}}
\end{array}} \right]
\end{equation}
The converted PDF is no longer Gaussian. In fact, the error volume has the shape of a contact lens \cite{Tian2009}.
\section{The Proposed Method}

\subsection{Unbiased Converted Measurement}
It is convenient for monostatic radar to find a multiplicative bias in measurement conversion from polar to Cartesian coordinates by taking the expectation of the converted measurements \cite{705921}. However, there is no closed form of biases for the bistatic case due to the nonlinearity in measurement conversion. Following a similar methodology as in \cite{Tian2009}, we utilize Taylor series expansion to find the conversion bias in bistatic radar.

First order Taylor series expansion of the conventional measurement conversion (\ref{E1}) produces no bias due to measurement noises being zero-mean. In this letter, we make the assumption that in the Taylor series expansion of (\ref{E1}), all terms of third order and above are negligible. For the target position along the x-axis, the second order Taylor series expansion of the conventional measurement conversion about the ground truth is
\begin{equation}
\begin{array}{l}
{x_m} = f\left( {{b_m},{\alpha _m}} \right) \approx \\
{\left. {f + {w_b}\frac{{\partial f}}{{\partial b}} + {w_\alpha }\frac{{\partial f}}{{\partial \alpha }} + \frac{1}{2}w_b^2\frac{{{\partial ^2}f}}{{\partial {b^2}}} + {w_b}{w_\alpha }\frac{{{\partial ^2}f}}{{\partial b\partial \alpha }} + \frac{1}{2}w_\alpha ^2\frac{{{\partial ^2}f}}{{\partial {\alpha ^2}}}} \right|_{\left( {\bar b,\bar \alpha } \right)}}
\end{array}
\end{equation}
Taking the expected value of the expansion gives us the expected value of the conventional converted measurement
\begin{equation}
E\left[ {{x_m}} \right] \approx {\left. {f + \frac{1}{2}\sigma _b^2\frac{{{\partial ^2}f}}{{\partial {b^2}}} + \frac{1}{2}\sigma _\alpha ^2\frac{{{\partial ^2}f}}{{\partial {\alpha ^2}}}} \right|_{\left( {\bar b,\bar \alpha } \right)}}
\end{equation}
which indicates that the conventional measurement conversion (\ref{E1}) is biased. Rather than use a multiplicative term for debiasing the conventional measurement conversion as in the monostatic radar case, we define the bias correction term
\begin{equation}
{c_x} \approx {\left. {\frac{1}{2}\sigma _b^2\frac{{{\partial ^2}f}}{{\partial {b^2}}} + \frac{1}{2}\sigma _\alpha ^2\frac{{{\partial ^2}f}}{{\partial {\alpha ^2}}}} \right|_{\left( {\bar b,\bar \alpha } \right)}}
\end{equation}
Subtracting the additive bias from the conventional converted measurement yields the unbiased converted measurement
\begin{equation}
x_m^{UCM} \buildrel \Delta \over = f\left( {{b_m},{\alpha _m}} \right) - {c_x}\left( {\bar b,\bar \alpha } \right)
\end{equation}
The calculation of the bias correction term requires the true target position, which is unavailable in practice. A feasible resolution to this problem is to evaluate the term at the measurement
\begin{equation}
\label{E2}
x_m^{UCM} \approx {\left. {f - \frac{1}{2}\sigma _b^2\frac{{{\partial ^2}f}}{{\partial {b^2}}} - \frac{1}{2}\sigma _\alpha ^2\frac{{{\partial ^2}f}}{{\partial {\alpha ^2}}}} \right|_{\left( {{b_m},{\alpha _m}} \right)}}
\end{equation}
Similarly, the bias correction term and the unbiased converted measurement for the target position along the y-axis are defined as follows, respectively
\begin{equation}
{c_y} \approx {\left. {\frac{1}{2}\sigma _b^2\frac{{{\partial ^2}g}}{{\partial {b^2}}} + \frac{1}{2}\sigma _\alpha ^2\frac{{{\partial ^2}g}}{{\partial {\alpha ^2}}}} \right|_{\left( {\bar b,\bar \alpha } \right)}}
\end{equation}
\begin{equation}
\label{E3}
\begin{array}{l}
y_m^{UCM} \buildrel \Delta \over = g\left( {{b_m},{\alpha _m}} \right) - {c_y}\left( {\bar b,\bar \alpha } \right)\\
 \approx {\left. {g - \frac{1}{2}\sigma _b^2\frac{{{\partial ^2}g}}{{\partial {b^2}}} - \frac{1}{2}\sigma _\alpha ^2\frac{{{\partial ^2}g}}{{\partial {\alpha ^2}}}} \right|_{\left( {{b_m},{\alpha _m}} \right)}}
\end{array}
\end{equation}

The CMKF requires the calculation of the converted measurement error covariance. Assuming that the expected value of (\ref{E2}) and (\ref{E3}) becomes the ground truth, and substituting the raw measurement for the true target position in practice, the entries in the corresponding converted measurement error covariance are
\begin{equation}
\label{E4}
R_{UCM}^{11} \approx {\left. \begin{array}{l}
\sigma _b^2{\left( {\frac{{\partial f}}{{\partial b}}} \right)^2} + \sigma _\alpha ^2{\left( {\frac{{\partial f}}{{\partial \alpha }}} \right)^2} + \frac{1}{2}\sigma _b^4{\left( {\frac{{{\partial ^2}f}}{{\partial {b^2}}}} \right)^2}\\
 + \frac{1}{2}\sigma _\alpha ^4{\left( {\frac{{{\partial ^2}f}}{{\partial {\alpha ^2}}}} \right)^2} + \sigma _b^2\sigma _\alpha ^2{\left( {\frac{{{\partial ^2}f}}{{\partial b\partial \alpha }}} \right)^2}
\end{array} \right|_{\left( {{b_m},{\alpha _m}} \right)}}
\end{equation}
\begin{equation}
\label{E5}
R_{UCM}^{22} \approx {\left. \begin{array}{l}
\sigma _b^2{\left( {\frac{{\partial g}}{{\partial b}}} \right)^2} + \sigma _\alpha ^2{\left( {\frac{{\partial g}}{{\partial \alpha }}} \right)^2} + \frac{1}{2}\sigma _b^4{\left( {\frac{{{\partial ^2}g}}{{\partial {b^2}}}} \right)^2}\\
 + \frac{1}{2}\sigma _\alpha ^4{\left( {\frac{{{\partial ^2}g}}{{\partial {\alpha ^2}}}} \right)^2} + \sigma _b^2\sigma _\alpha ^2{\left( {\frac{{{\partial ^2}g}}{{\partial b\partial \alpha }}} \right)^2}
\end{array} \right|_{\left( {{b_m},{\alpha _m}} \right)}}
\end{equation}
\begin{equation}
\label{E6}
R_{UCM}^{12} \approx {\left. \begin{array}{l}
\sigma _b^2\frac{{\partial f}}{{\partial b}}\frac{{\partial g}}{{\partial b}} + \sigma _\alpha ^2\frac{{\partial f}}{{\partial \alpha }}\frac{{\partial g}}{{\partial \alpha }} + \frac{1}{2}\sigma _b^4\frac{{{\partial ^2}f}}{{\partial {b^2}}}\frac{{{\partial ^2}g}}{{\partial {b^2}}}\\
 + \frac{1}{2}\sigma _\alpha ^4\frac{{{\partial ^2}f}}{{\partial {\alpha ^2}}}\frac{{{\partial ^2}g}}{{\partial {\alpha ^2}}} + \sigma _b^2\sigma _\alpha ^2\frac{{{\partial ^2}f}}{{\partial b\partial \alpha }}\frac{{{\partial ^2}g}}{{\partial b\partial \alpha }}
\end{array} \right|_{\left( {{b_m},{\alpha _m}} \right)}}
\end{equation}

Equations (\ref{E2}), (\ref{E3}), and (\ref{E4}-\ref{E6}) constitute the Unbiased Converted Measurement (UCM) for bistatic measurement conversion.

\subsection{Decorrelated Unbiased Converted Measurement}
Evaluating the converted measurement error covariance using the raw measurement results in the correlation between this covariance estimate and the measurement error. Consequently, the filter gain of the CMKF becomes dependent on the measurement noise, which leads to an estimation bias when the converted measurement is used in tracking \cite{Bordonaro2014,10.1117/12.895484}. In the monostatic case, to decorrelate the converted measurement error covariance from the measurement noise, either the prediction at the current step \cite{Spitzmiller2010} or the measurement at the previous step \cite{10.1117/12.831218} is utilized to evaluate the covariance. Generally, the prediction is more accurate than the previous measurement.

For the bistatic case, the prediction of the bistatic range and bearing at the current step is
\begin{equation}
\begin{array}{l}
{b_t} = \bar b + {w_{{b_t}}}\\
{\alpha _t} = \bar \alpha  + {w_{{\alpha _t}}}
\end{array}
\end{equation}
where ${w_{{b_t}}}$ and ${w_{{\alpha _t}}}$ are independent zero-mean Gaussian white noise with variance $\sigma _{{b_t}}^2$ and $\sigma _{{\alpha _t}}^2$, respectively, and uncorrelated with the measurement noise. Based on this assumption, the second order Taylor series expansion of $x_m^{UCM} - E\left[ {x_m^{UCM}} \right]$ and $y_m^{UCM} - E\left[ {y_m^{UCM}} \right]$ about the prediction yields the decorrelated converted measurement error covariance
\begin{equation}
\label{E7}
\begin{array}{l}
R_{DUCM}^{11} \approx \\
{\left. \begin{array}{l}
\sigma _b^2{\left( {\frac{{\partial f}}{{\partial b}}} \right)^2} + \sigma _\alpha ^2{\left( {\frac{{\partial f}}{{\partial \alpha }}} \right)^2} + \frac{1}{2}\sigma _b^4{\left( {\frac{{{\partial ^2}f}}{{\partial {b^2}}}} \right)^2}\\
 + \frac{1}{2}\sigma _\alpha ^4{\left( {\frac{{{\partial ^2}f}}{{\partial {\alpha ^2}}}} \right)^2} + \sigma _b^2\sigma _{{b_t}}^2{\left( {\frac{{{\partial ^2}f}}{{\partial {b^2}}}} \right)^2} + \sigma _\alpha ^2\sigma _{{\alpha _t}}^2{\left( {\frac{{{\partial ^2}f}}{{\partial {\alpha ^2}}}} \right)^2}\\
 + \left( {\sigma _b^2\sigma _\alpha ^2 + \sigma _b^2\sigma _{{\alpha _t}}^2 + \sigma _\alpha ^2\sigma _{{b_t}}^2} \right){\left( {\frac{{{\partial ^2}f}}{{\partial b\partial \alpha }}} \right)^2}
\end{array} \right|_{\left( {{b_t},{\alpha _t}} \right)}}
\end{array}
\end{equation}
\begin{equation}
\label{E8}
\begin{array}{l}
R_{DUCM}^{22} \approx \\
{\left. \begin{array}{l}
\sigma _b^2{\left( {\frac{{\partial g}}{{\partial b}}} \right)^2} + \sigma _\alpha ^2{\left( {\frac{{\partial g}}{{\partial \alpha }}} \right)^2} + \frac{1}{2}\sigma _b^4{\left( {\frac{{{\partial ^2}g}}{{\partial {b^2}}}} \right)^2}\\
 + \frac{1}{2}\sigma _\alpha ^4{\left( {\frac{{{\partial ^2}g}}{{\partial {\alpha ^2}}}} \right)^2} + \sigma _b^2\sigma _{{b_t}}^2{\left( {\frac{{{\partial ^2}g}}{{\partial {b^2}}}} \right)^2} + \sigma _\alpha ^2\sigma _{{\alpha _t}}^2{\left( {\frac{{{\partial ^2}g}}{{\partial {\alpha ^2}}}} \right)^2}\\
 + \left( {\sigma _b^2\sigma _\alpha ^2 + \sigma _b^2\sigma _{{\alpha _t}}^2 + \sigma _\alpha ^2\sigma _{{b_t}}^2} \right){\left( {\frac{{{\partial ^2}g}}{{\partial b\partial \alpha }}} \right)^2}
\end{array} \right|_{\left( {{b_t},{\alpha _t}} \right)}}
\end{array}
\end{equation}
\begin{equation}
\label{E9}
\begin{array}{l}
R_{DUCM}^{12} \approx \\
{\left. \begin{array}{l}
\sigma _b^2\frac{{\partial f}}{{\partial b}}\frac{{\partial g}}{{\partial b}} + \sigma _\alpha ^2\frac{{\partial f}}{{\partial \alpha }}\frac{{\partial g}}{{\partial \alpha }} + \frac{1}{2}\sigma _b^4\frac{{{\partial ^2}f}}{{\partial {b^2}}}\frac{{{\partial ^2}g}}{{\partial {b^2}}}\\
 + \frac{1}{2}\sigma _\alpha ^4\frac{{{\partial ^2}f}}{{\partial {\alpha ^2}}}\frac{{{\partial ^2}g}}{{\partial {\alpha ^2}}} + \sigma _b^2\sigma _{{b_t}}^2\frac{{{\partial ^2}f}}{{\partial {b^2}}}\frac{{{\partial ^2}g}}{{\partial {b^2}}} + \sigma _\alpha ^2\sigma _{{\alpha _t}}^2\frac{{{\partial ^2}f}}{{\partial {\alpha ^2}}}\frac{{{\partial ^2}g}}{{\partial {\alpha ^2}}}\\
 + \left( {\sigma _b^2\sigma _\alpha ^2 + \sigma _b^2\sigma _{{\alpha _t}}^2 + \sigma _\alpha ^2\sigma _{{b_t}}^2} \right)\frac{{{\partial ^2}f}}{{\partial b\partial \alpha }}\frac{{{\partial ^2}g}}{{\partial b\partial \alpha }}
\end{array} \right|_{\left( {{b_t},{\alpha _t}} \right)}}
\end{array}
\end{equation}

In practice, the prediction is usually estimated in Cartesian coordinates with the predicted position ${x_t}$ and ${y_t}$ and its associated covariance ${P_t}$. The predicted bistatic range and bearing are
\begin{equation}
{b_t} = \varphi \left( {{x_t},{y_t}} \right)
\end{equation}
\begin{equation}
{\alpha _t} = \gamma \left( {{x_t},{y_t}} \right)
\end{equation}
Their associated variances are approximated by linearizing the track's covariance
\begin{equation}
\sigma _{{b_t}}^2 = {\left. {\left[ {\begin{array}{*{20}{c}}
{\frac{{\partial \varphi }}{{\partial x}}}&{\frac{{\partial \varphi }}{{\partial y}}}
\end{array}} \right]{P_t}\left[ {\begin{array}{*{20}{c}}
{\frac{{\partial \varphi }}{{\partial x}}}\\
{\frac{{\partial \varphi }}{{\partial y}}}
\end{array}} \right]} \right|_{\left( {{x_t},{y_t}} \right)}}
\end{equation}
\begin{equation}
\sigma _{{\alpha _t}}^2 = {\left. {\left[ {\begin{array}{*{20}{c}}
{\frac{{\partial \gamma }}{{\partial x}}}&{\frac{{\partial \gamma }}{{\partial y}}}
\end{array}} \right]{P_t}\left[ {\begin{array}{*{20}{c}}
{\frac{{\partial \gamma }}{{\partial x}}}\\
{\frac{{\partial \gamma }}{{\partial y}}}
\end{array}} \right]} \right|_{\left( {{x_t},{y_t}} \right)}}
\end{equation}
which ignore the correlation between the predicted bistatic range and bearing.

Equations (\ref{E2}), (\ref{E3}), and (\ref{E7}-\ref{E9}) constitute the Decorrelated Unbiased Converted Measurement (DUCM) for bistatic measurement conversion.

\section{Evaluation}
The evaluation of the proposed technique includes the analysis of the static conversion and the use of the converted measurement in a dynamic tracking scenario.
\subsection{Analysis of Static Conversion}
An ideal measurement conversion would be unbiased and consistent. Conversion bias is defined as the difference between the expected value of the converted measurement and the truth.

Fig. 2 shows the results of Monte Carlo simulation for various bearing angles, verifying that UCM/DUCM is unbiased.

A common measure of consistency is the normalized estimation error squared (NEES) \cite{Crouse2014}
\begin{equation}
NEES = \frac{1}{{{d^x}}}E\left[ {{{\left( {{t_m} - \bar t} \right)}^T}R_{xy}^{ - 1}\left( {{t_m} - \bar t} \right)} \right]
\end{equation}
where ${d^x}$ is the dimensionality of the state, $\bar t$ is the ground truth of target position in Cartesian coordinates, and $R_{xy}$ is the converted measurement error covariance. The NEES of a consistent conversion should be close to $1$.

The conventional conversion method, defined as (\ref{E1}) with the converted measurement error covariance $J{R_{b\alpha }}{J^T}$, is adopted as the baseline. Fig. 3 depicts NEES for conventional, UCM, and DUCM conversion methods based on 10 000 Monte Carlo runs. An upper and lower bounds for a $99\%$ confidence region of the NEES are approximately $0.9744$ and $1.0259$. DUCM remains consistent overall, while the conventional method and UCM method exhibit inconsistencies in some case, such as a long bistatic range, bearings near ${0^ \circ }$, high range resolution, or low bearing resolution.
\begin{figure}
\centerline{\includegraphics[width=8.6cm]{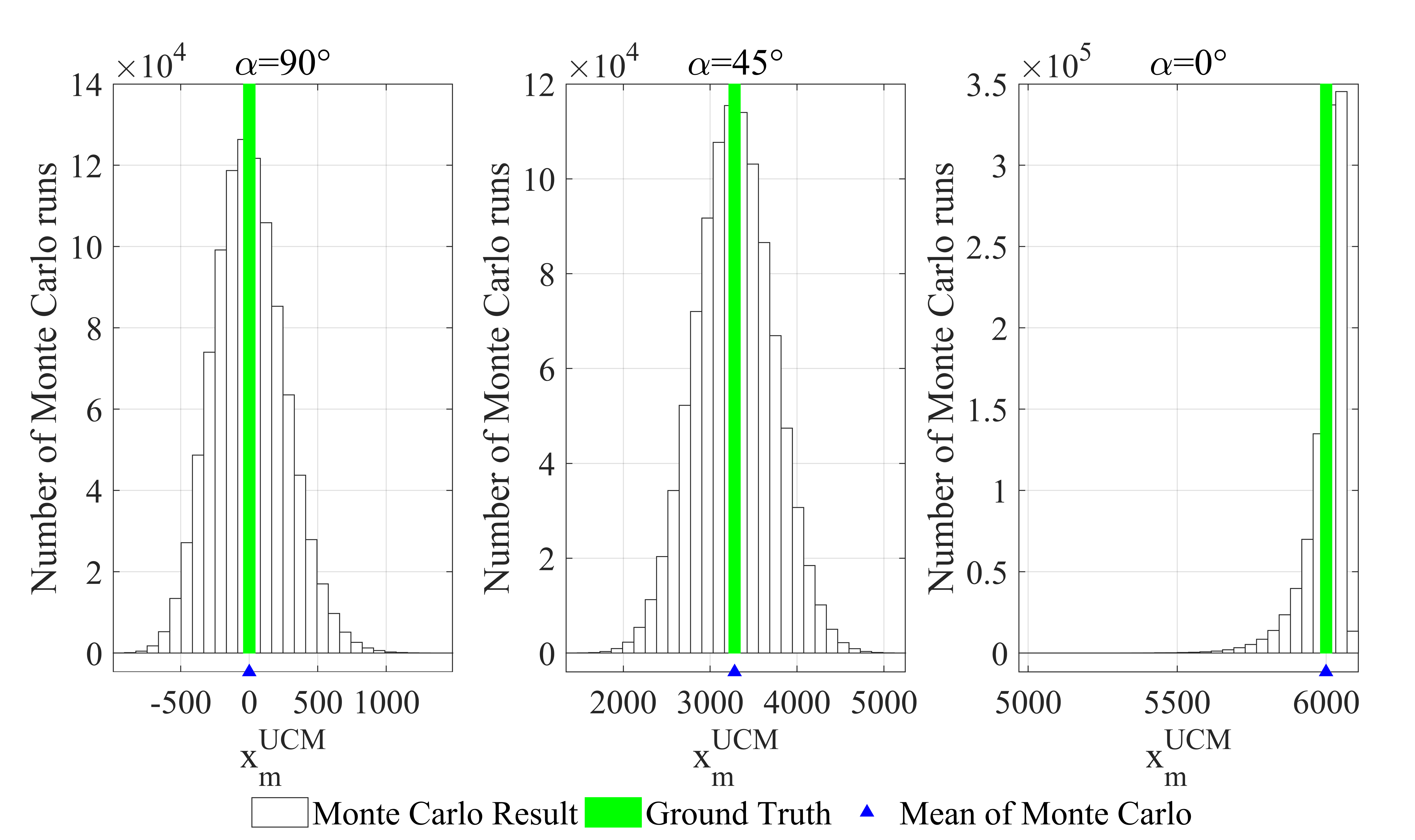}}
\caption{Histogram of 1 000 000 Monte Carlo runs using a bistatic range of $8000m$ and bearing of ${0^ \circ }$ to ${90^ \circ }$ with ${\sigma _b} = 30m$, ${\sigma _\alpha } = {5^ \circ }$, and $L = 4000m$. Mean value of Monte Carlo result and ground truth are plotted for comparison. In all cases, UCM was unbiased.}
\end{figure}
\begin{figure}
\centering
\subfigure[]{\includegraphics[width=4.3cm]{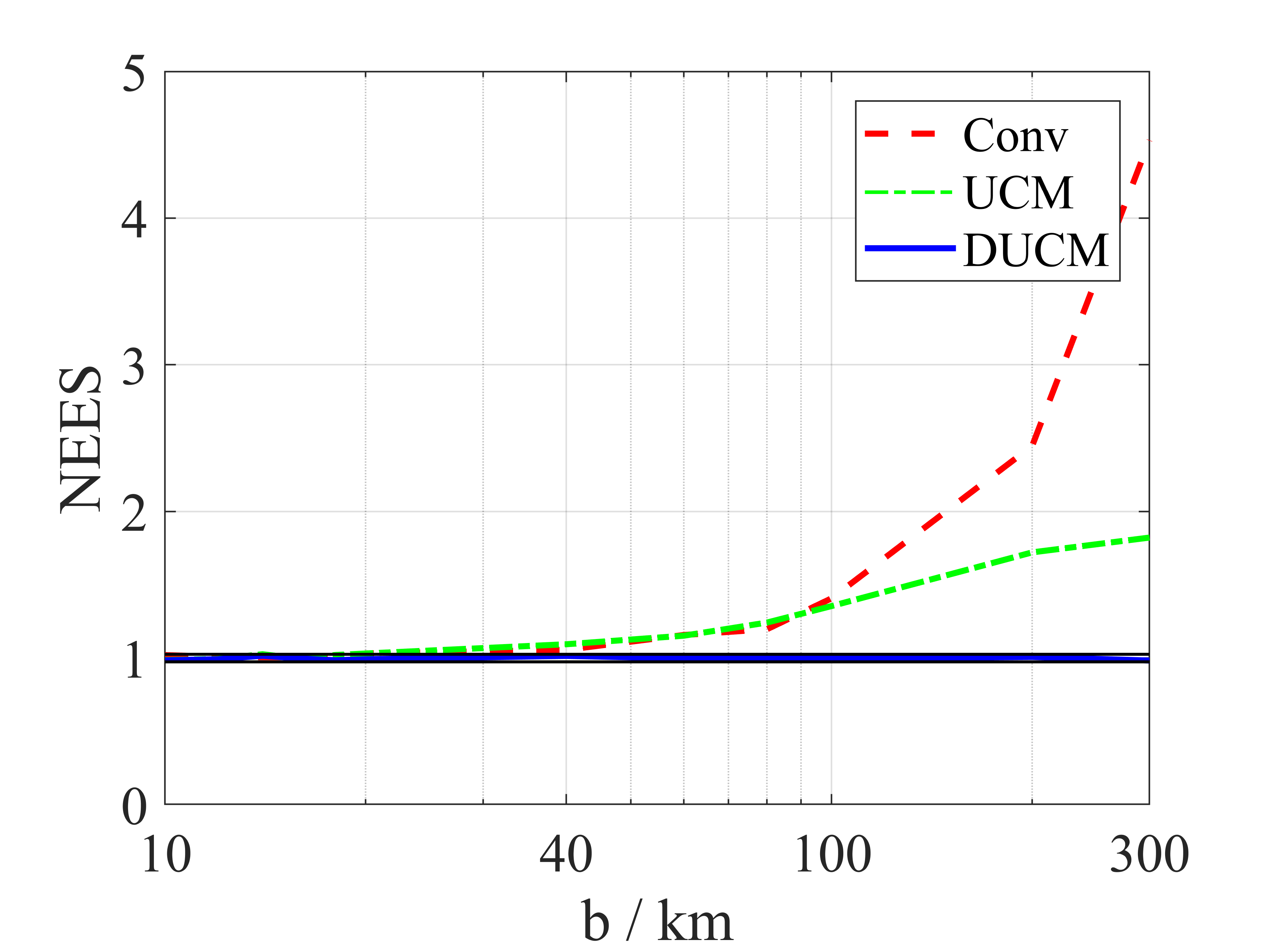}}
\subfigure[]{\includegraphics[width=4.3cm]{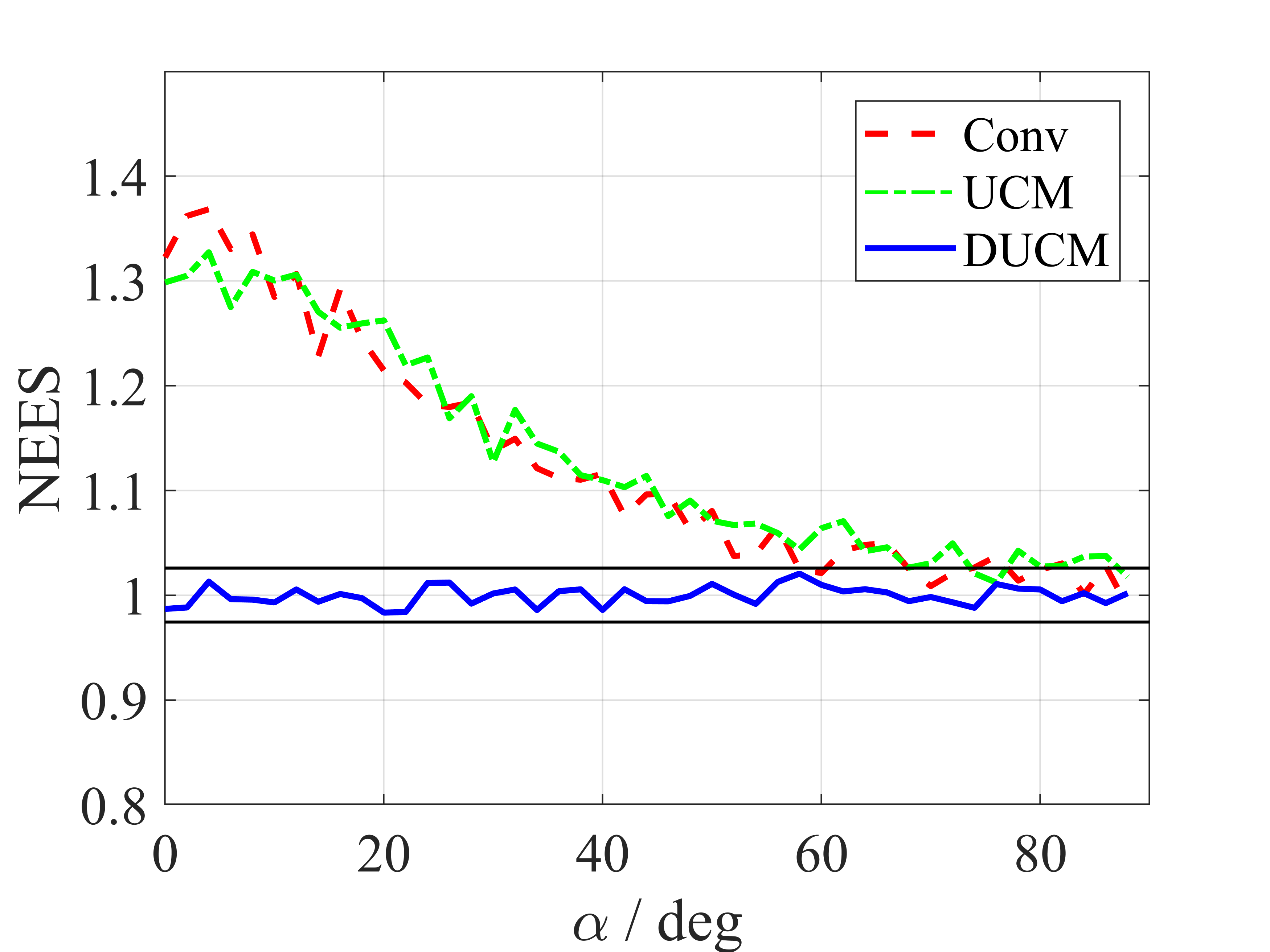}}
\subfigure[]{\includegraphics[width=4.3cm]{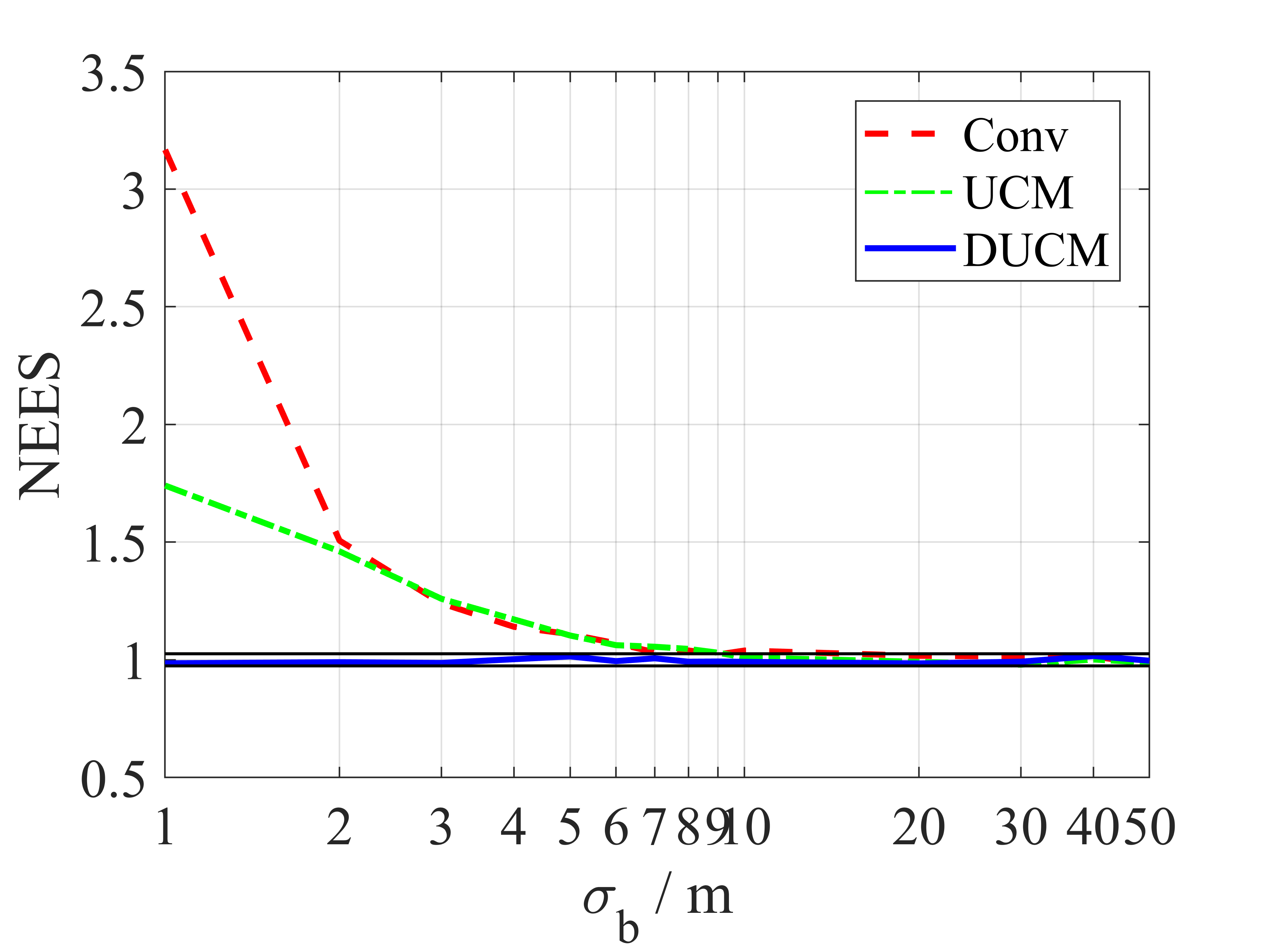}}
\subfigure[]{\includegraphics[width=4.3cm]{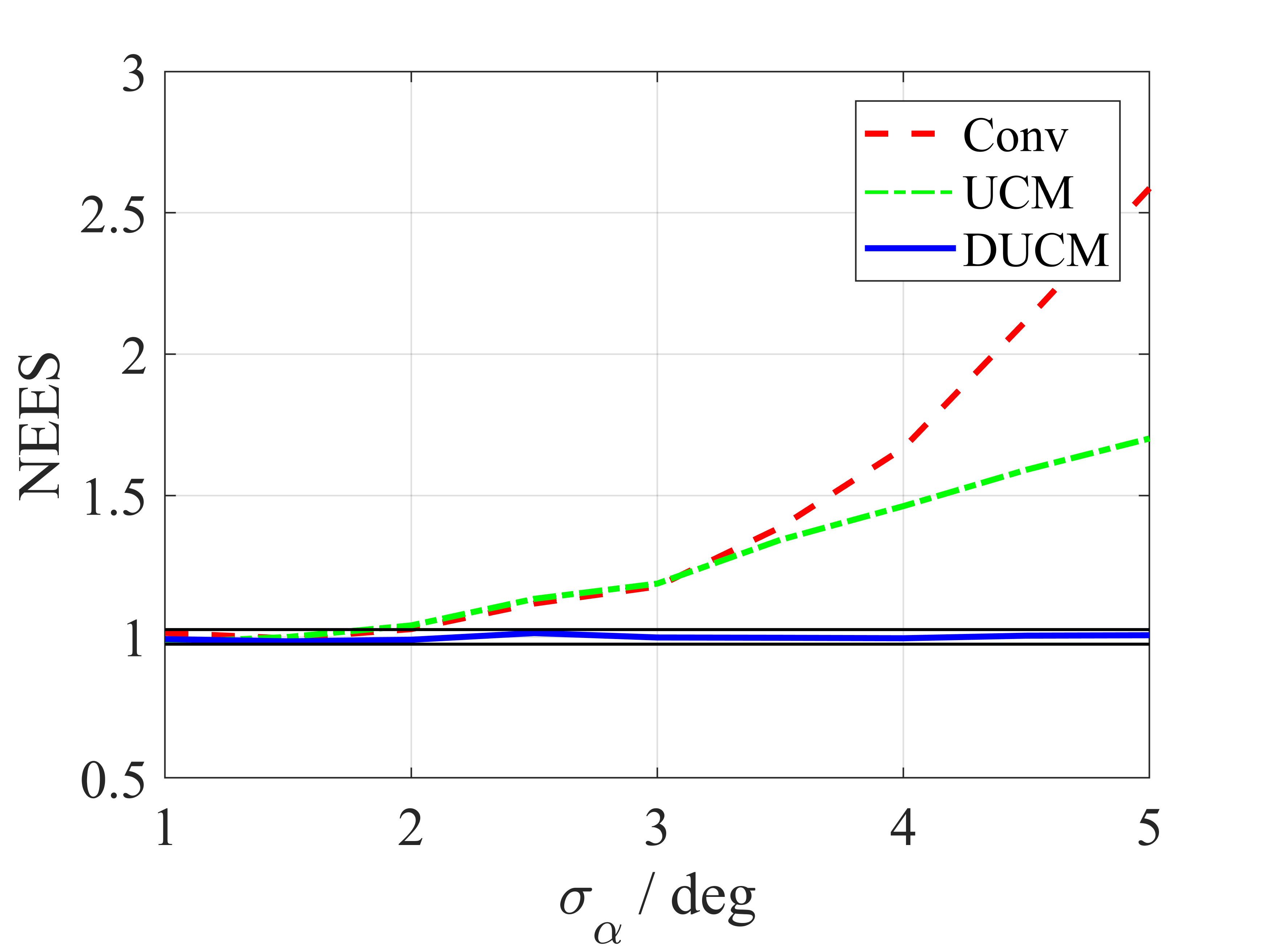}}
\caption{NEES for conventional, UCM, and DUCM conversion methods based on 10 000 Monte Carlo runs with $L = 4000m$. For the calculation of DUCM, the predicted position ${x_t}$ and ${y_t}$ are set to the ground truth state corrupted by normally distributed noise with ${P_t} = \sigma _b^2 \times \left[ {1,0.1;0.1,1} \right]$. (a) NEES versus $b$ with ${\sigma _b} = 30m$, ${\sigma _\alpha } = {1^ \circ }$, and the target is always on the perpendicular bisector of the baseline. (b) NEES versus $\alpha$ with ${\sigma _b} = 30m$, ${\sigma _\alpha } = {2^ \circ }$, and $b = 8000m$. (c) NEES versus ${\sigma _b}$ with ${\sigma _\alpha } = {1^ \circ }$, $b = 8000m$, and $\alpha  = {60^ \circ }$. (d) NEES versus ${\sigma _\alpha }$ with ${\sigma _b} = 30m$, $b = 8000m$, and $\alpha  = {60^ \circ }$. Plots include chi-square 0.99 probability bounds.}
\end{figure}
\subsection{Application to Converted Measurement Kalman Filter}
This subsection evaluates aforementioned three converted measurement Kalman filters for bistatic radar tracking. In the simulations, the baseline $L=4000m$, and the sampling period $T=1s$. The target starts from ${\left[ {8000m,8000m} \right]^T}$ in Cartesian coordinates with initial velocity $10m{s^{ - 1}}$ and random direction distributed uniformly from $0$ to $2\pi$. The state of the target is defined in Cartesian coordinates $X = {\left[ {x,\dot x,y,\dot y} \right]^T}$, and a simple linear motion model is used, the discretized continuous white noise acceleration (DCWNA) model. The covariance of the discretized process noise is
\begin{equation}
\left[ {\begin{array}{*{20}{c}}
{{\rm{0}}{\rm{.0625}}}&{{\rm{0}}{\rm{.125}}}&0&0\\
{{\rm{0}}{\rm{.125}}}&{{\rm{0}}{\rm{.25}}}&0&0\\
0&0&{{\rm{0}}{\rm{.0625}}}&{{\rm{0}}{\rm{.125}}}\\
0&0&{{\rm{0}}{\rm{.125}}}&{{\rm{0}}{\rm{.25}}}
\end{array}} \right]
\end{equation}
and measurement noise standard deviations are ${\sigma _b} = 10m$, ${\sigma _\alpha } = {2^ \circ }$. All three filters are initialized with the position of initial measurement, the velocity of $0$, and the covariance of $diag\left( {\left[ {100,100,100,100} \right]} \right)$. Estimation accuracy of position is evaluated by the root mean square error (RMSE)
\begin{equation}
RMS{E_p} = \sqrt {E\left[ {{{\left\| {{X_p} - {{\hat X}_p}} \right\|}^2}} \right]}
\end{equation}
where the ground truth and the estimate of the position component of target state are ${X_p}$ and ${\hat X_p}$, respectively. Velocity RMSE has a similar definition. The NEES is used to evaluate covariance consistency
\begin{equation}
NEES = \frac{1}{{{d_x}}}E\left[ {{{\left( {X - \hat X} \right)}^T}{P^{ - 1}}\left( {X - \hat X} \right)} \right]
\end{equation}
where $\hat X$ is the estimate of target state, and $P$ is the error covariance. Fig. 4 depicts 5000-run Monte Carlo evaluation of CMKF using conventional, UCM and DUCM conversion. Fig. 4(a) and (b) demonstrate that the DUCM filter has the best RMSE performance for position and velocity, respectively, while the UCM filter outperforms the conventional filter. Fig. 4(c) shows that the DUCM filter exhibits the improved consistency with a NEES close to one after scan $110$, whereas the conventional filter and the UCM filter are clearly inconsistent. It should be mentioned that the DUCM filter converges most slowly due to the dependence on the prediction.
\begin{figure}
\centering
\subfigure[]{\includegraphics[width=4.3cm]{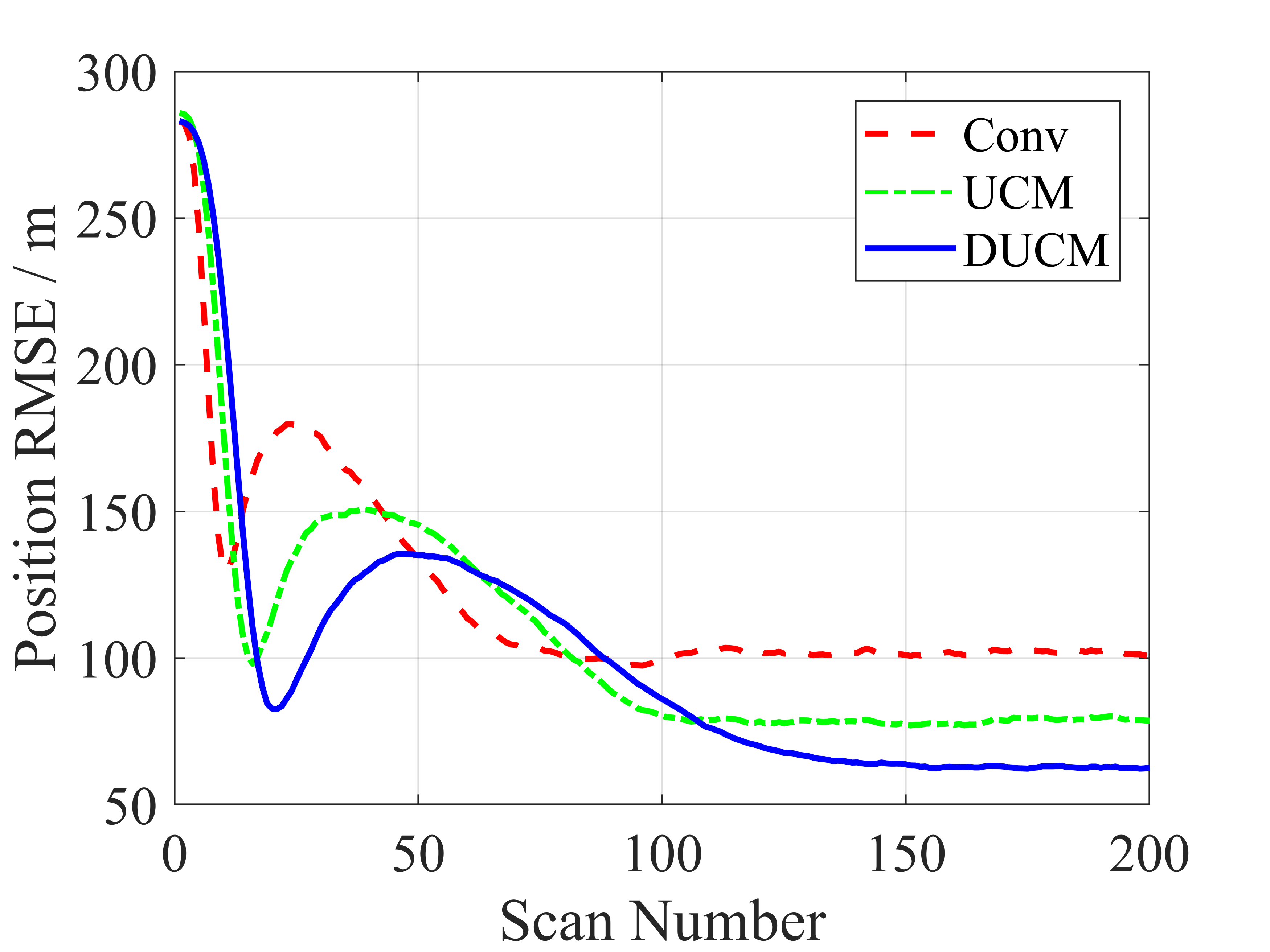}}
\subfigure[]{\includegraphics[width=4.3cm]{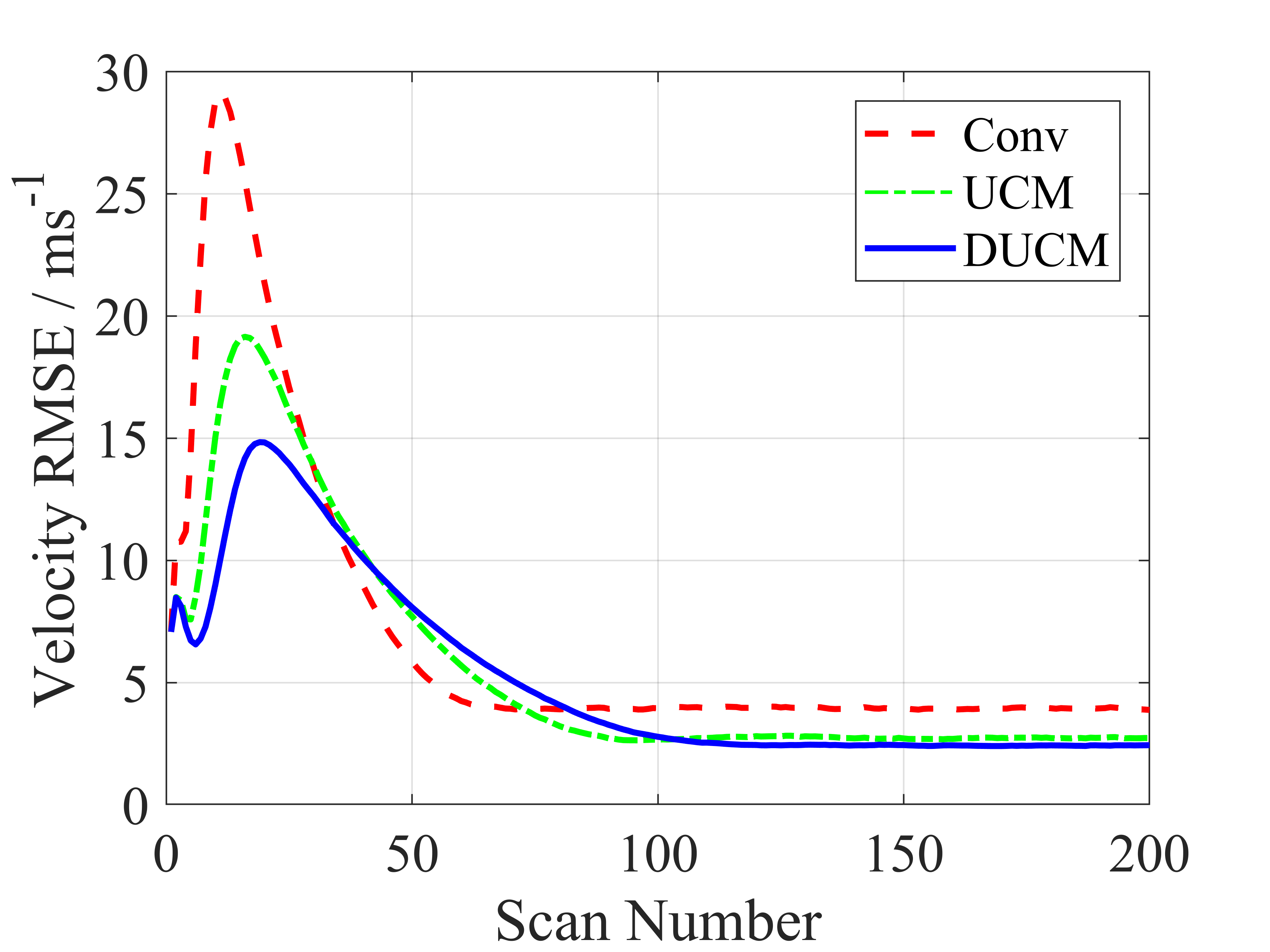}}
\subfigure[]{\includegraphics[width=4.3cm]{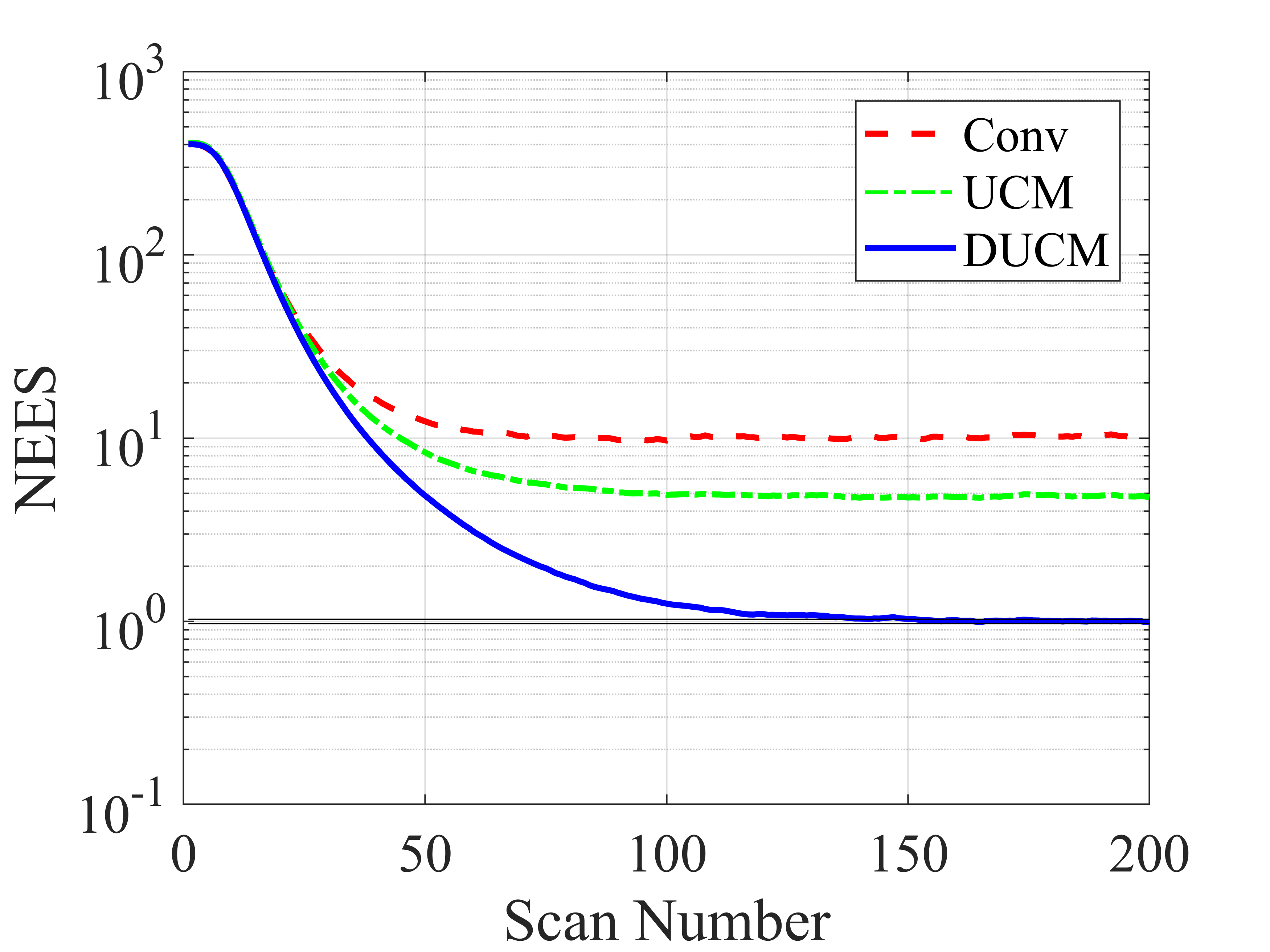}}
\caption{5000-run Monte Carlo evaluation of CMKF using conventional, UCM and DUCM conversion. (a) Root mean square error for position estimate. (b) Root mean square error for velocity estimate. (c) NEES including chi-square 0.99 probability bounds.}
\end{figure}
\section{Conclusion}
The converted measurement Kalman filter is a category of approach to deal with the nonlinearity of measurements. Based on Taylor series expansion, this letter eliminates the conversion bias in bistatic measurement conversion by the UCM, and proposes a decorrelated version of the UCM technique (DUCM) to avoid correlation of the converted measurement error covariance estimate and the measurement noise. Simulations of static conversion verify the unbiasedness and the consistency of the DUCM conversion, and simulations of dynamic tracking demonstrate that the DUCM filter has superior estimation accuracy and covariance consistency than the conventional CMKF and the UCM filter.
\bibliography{CoordinateConversion}
\end{document}